\documentclass{iaus}
\usepackage{graphics}

\title[Nebular emission lines in IRAS\,17347--3139] 
{Nebular emission lines in IRAS\,17347--3139}

\author[Jim\'enez-Esteban et al.]   
{F.M. Jim\'enez-Esteban$^1$, 
J.V. Perea-Calder\'on$^2$, \break O. Su\'arez$^3$, M. Bobrowsky$^4$, 
P. Garc\'{\i}a-Lario$^{5}$}

\affiliation{$^1$Hamburger Sternwarte, Hamburg, Germany \break email: francisco.jimenez-esteban@hs.uni-hamburg.de\\[\affilskip]
$^2$European Space Astronomy Centre (ESAC), Madrid, Spain\\
$^3$Laboratorio de Astrof\'{\i}sica Espacial y F\'{\i}sica Fundamental-INTA, Madrid, Spain\\
$^4$Computer Science Corporation/Space Telescope Science Institute, Baltimore
MD,  USA\\
$^5$ESAC, 
    Research and Scientific Support Deparment of ESA, Madrid, Spain}
\pubyear{2006}
\volume{234}  
\pagerange{}
\date{?? and in revised form ??}
\jname{Planetary Nebula in Our Galaxy and Beyond}
\editors{}
\begin{document}

\maketitle

\begin{abstract}
  We report the detection of nebular emission lines in the optical and
  mid-infrared spectra of IRAS\,17347--3139, a heavily obscured OH/IR
  star which may be rapidly evolving from the AGB to the PN stage. The
  presence of emission lines is interpreted as a clear indication that
  the ionization of its circumstellar envelope has already
  started. This source belongs to the rare class of objects known as
  `OHPNe' displaying both OH maser and radio continuum
  emission. However, unlike the rest of stars in this class, prominent
  C-rich dust features are detected in its mid-infrared spectrum,
  which makes the analysis of this star particularly interesting.
\end{abstract}

\firstsection 

\section{Introduction}

As low- and intermediate-mass (0.8\,$-$\,8\,M$_{\odot}$) stars evolve
along the Asymptotic Giant Branch (AGB) in their process to become
Planetary Nebulae (PNe) they experience strong mass loss (up to
10$^{-4}$\,M$_{\odot}$\,yr$^{-1}$) which results in the formation of gas
and dust circumstellar envelopes (CSEs) which sometimes may become
thick enough to completely obscure the light coming from the central
star in the optical at the end of the AGB phase, when the mass loss
rate is maximum.
In particular, O-rich stars evolving from the AGB to the PN stage develop a 
strong and characteristic double-peaked OH maser emission at 1612 MHz at the
end of their AGB lifetime which later disappears when the mass loss
drops below a certain level as the star enters the post-AGB stage. 
An interesting and peculiar class of sources recently
discovered is the one formed by the so-called `OHPNe'
\cite{Zijlstra89}. OHPNe are heavily obscured O-rich transition
sources which show both this OH maser emission as well
as radio continuum emission, indicative of ongoing ionization in their
circumstellar envelope. They have been proposed to be the `missing
link' between high-mass O-rich AGB stars experiencing hot bottom burning 
and Type I PNe.  These massive objects may evolve so fast that when 
the central star reaches the necessary
 temperature to ionize the envelope, this is still thick enough to
 prevent the detection of the PN in the optical. 
 These massive AGB stars may never be detectable in the optical as classical 
PNe.   Rather, they would evolve as `infrared PNe'.
 Unfortunately, the number of OHPNe known is very small (a dozen
sources) to derive firm conclusions on their nature and evolutionary stage. 

 Here we report on the identification of nebular emission lines in 
IRAS\,17347--3139 (= GLMP\,591), another member of
this class showing some peculiar observational properties.


The optical observations here presented were carried out in June 2003
using EFOSC2 while the mid-infrared observations were carried out in July 2003 
with TIMMI2, both instruments installed at the ESO 3.6\,m telescope in  
La Silla (Chile). 
The infrared spectrum covering the spectral range from 2 to 45\,$\mu$m was
retrieved from the ISO Data Archive, and it was taken with the Short 
Wavelength Spectrometer in the SWS01 observing mode.

\section{Results}



The optical spectrum of IRAS\,17347--3139 is shown in Figure 1 (left panel),
where we can only see a very red continuum together with a strong [S III]
9069\,\AA~line in emission. In addition, Br$_{\alpha}$ emission at
4.05\,$\mu$m and a prominent [Ne II] emission line at 12.81\,$\mu$m are clearly
identified in the ISO SWS01 spectrum. The detection of [Ne II] was later
confirmed in our mid-infrared observations with TIMMI2 (see Figure 1 right
panel). This suggests that the ionization of the circumstellar shell has
already started.

 The detection of OH \cite{Zijlstra89} and H$_{2}$O maser emission
\cite{Gregorio-Monsalvo04} is a clear signature of the O-rich nature of the
CSE surrounding this source. Unexpectedly, however, the mid-infrared spectrum
shows the presence of broad and prominent solid state emission features at
6.2, 7,7, 8.6 and 11.3\,$\mu$m usually attributed to Policyclic Aromatic
Hydrocarbons (PAHs) in addition to a broad plateau of emission from 11 to
15\,$\mu$m, characteristic also of hydrogenated PAHs (see Figure 1; right panel),
which suggest that there is also a substantial amount of C-rich dust in the
shell.

This dual-dust chemistry has only been observed before in a few other 
transition sources and 
several hypothesis have been proposed to explain the phenomenon.
Among them: a) a very recent thermal pulse which may have efficiently dredged
up C-rich processed material to the stellar surface (Waters et al. 1998);
 b) a long-lived
circumbinary disk which has preserved the O-rich material from destruction
(Molster et al. 1999);
and c) a result of the deactivation of hot bottom burning at the end of the AGB
as a consequence of the strong mass loss  (Frost et al. 1998). 
In order to determine which of the above hypothesis is correct additional 
observations are needed
which should be able to reveal the relative distribution of O-rich and C-rich
material within the circumstellar envelope.

\begin{figure}
\centering
\resizebox{6.7cm}{!}{\includegraphics{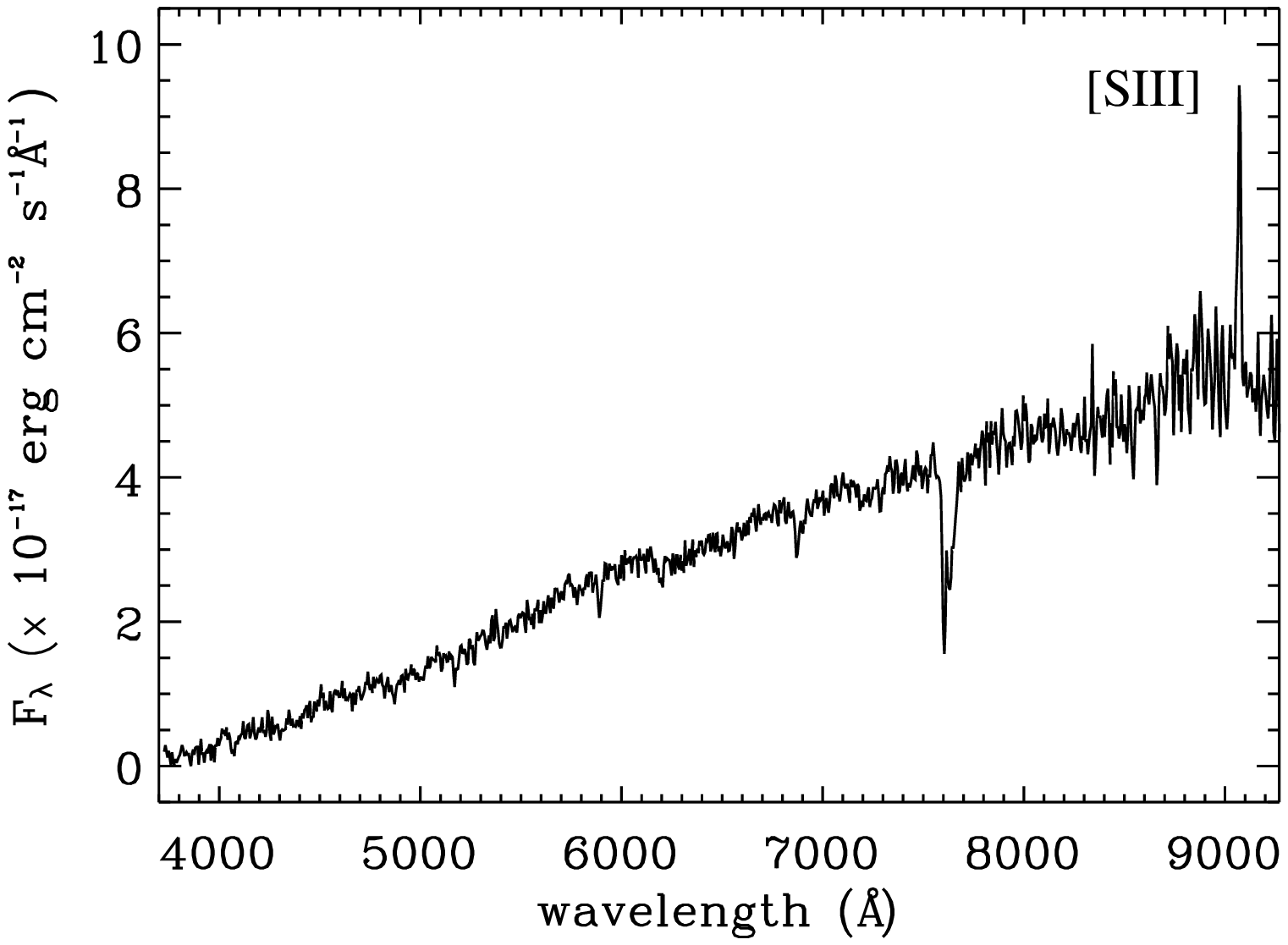} }
\resizebox{6.7cm}{!}{\includegraphics{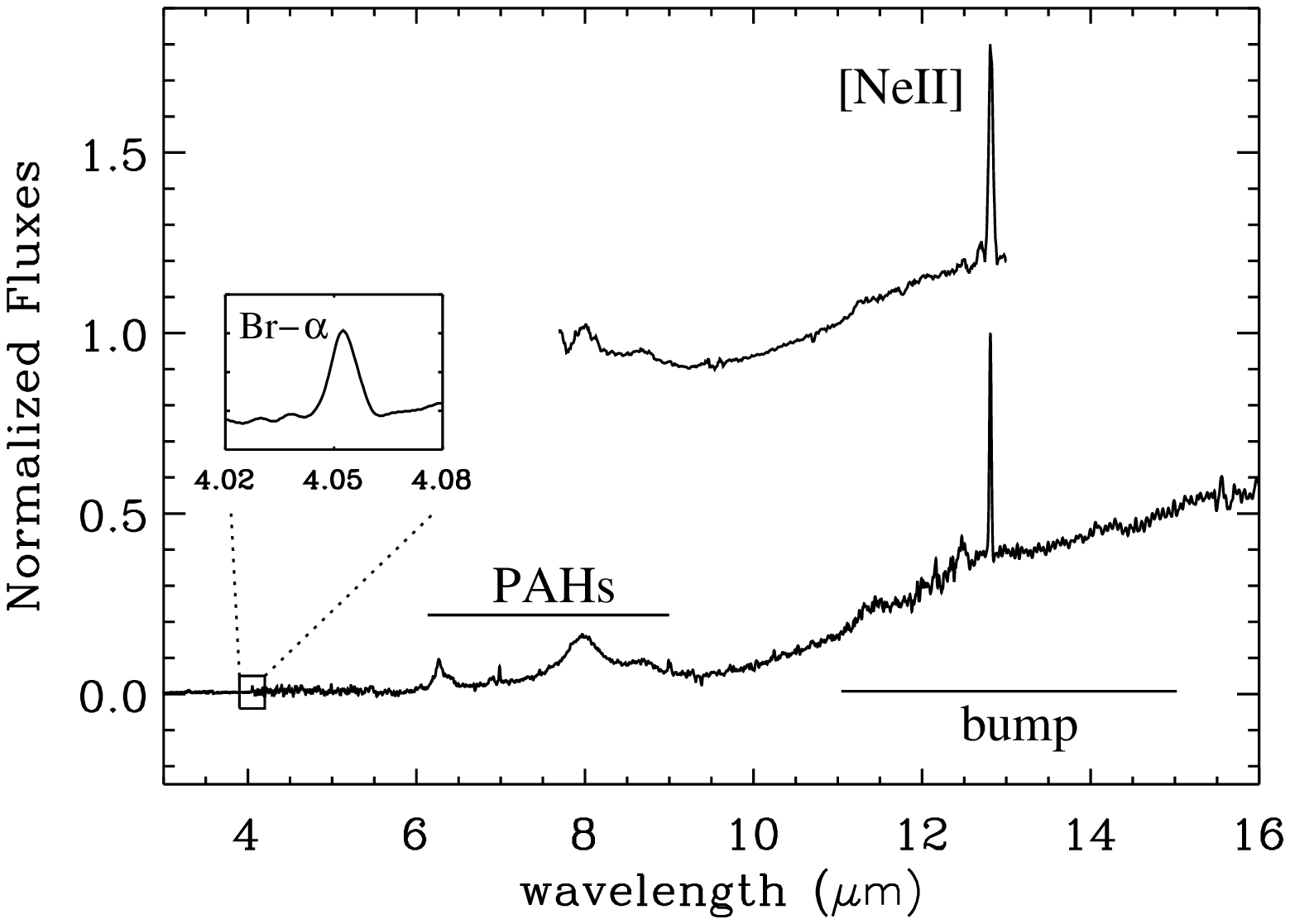} }
\caption[]{Left: The optical spectrum obtained with
  EFOSC2 at the ESO 3.6\,m telescope, where the prominent [S III]
  9068.6\,\AA~emission is indicated. Right: ISO-SWS spectrum (bottom) showing
  Br$_{\alpha}$ (4.05\,$\mu$m) and [Ne II] 12.81\,$\mu$m emission lines. PAH
  emission features at 6.2, 7.7 and 8.6\,$\mu$m are also detected together with
  a broad plateau of hydrogenated PAHs extending from 11 to 15\,$\mu$m. The
  detection of these features was confirmed with TIMMI2 (top)}
\end{figure}


\end{document}